\documentclass[preprint,3p]{elsarticle}
\usepackage{amssymb}
\usepackage{amsthm}
\usepackage{dcolumn}
\usepackage[mathlines]{lineno}
\usepackage{graphicx}
\usepackage{units}
\usepackage{url}
\usepackage{amsmath}
\usepackage{amsfonts}
\usepackage{bm}
\usepackage{textcomp}
\usepackage{subfigure}
\usepackage{multicol}
\usepackage{verbatim}
\usepackage{rotating}
\usepackage[colorlinks,linkcolor=blue,anchorcolor=blue,citecolor=blue]{hyperref}
\usepackage{float}
\usepackage{epsfig}
\usepackage{dcolumn}
\usepackage{bm}
\usepackage{color}
\usepackage{pstricks}
\usepackage{pst-node}
\usepackage{times}
\usepackage{indentfirst}
\usepackage[english]{babel}
\usepackage{ulem}
\addto{\captionsenglish}{%

}
\journal{Physics Letters B}
\makeatletter

\newenvironment{tablehere}
  {\def\@captype{table}}
  {}

\newenvironment{figurehere}
  {\def\@captype{figure}}
  {}
\makeatother

\def \DDbar   {D^{0}\bar{D^{0}}}
\def \dpdm    {D^{+}D^{-}}
\def \dz      {D^{0}}
\def \dzbar   {\bar{D^{0}}}
\def \de      {\Delta E}
\def \mbc     {M_{\rm BC}}
\def \kpipiz  {K^{-}\pi^{+}\pi^{0}}

\def \pipipi    {\pi^{0}\pi^{0}\pi^{0}}
\def \pipieta   {\pi^0\pi^0\eta}
\def \pietaeta  {\pi^{0}\eta\eta}
\def \etaetaeta {\eta\eta\eta}

\def \ks    {K_{S}^{0}}
\def \az    {a_{0}(980)^{0}}
\def \qqbar {q\bar{q}}
\def \dbar  {\bar{D}^0}

\def \ee   {e^+e^-}
\def \gev  {\mbox{GeV}}

\def \gevcc{\mbox{GeV/$c^2$}}

\def \mevcc{\mbox{MeV/$c^2$}}
\def \ifb  {\mbox{fb$^{-1}$}}
\def \ipb  {\mbox{pb$^{-1}$}}

\def \BR   {\mathcal{B}}

\def \piz  {\pi^0}
\def \pip  {\pi^+}
\def \pim  {\pi^-}

\begin{document}
\begin{frontmatter}
\title{{\bf
\boldmath Measurement of Singly Cabibbo-Suppressed Decays $\dz\to\pipipi$, $\pipieta$, $\pietaeta$ and $\etaetaeta$}}
\author{
M.~Ablikim$^{1}$, M.~N.~Achasov$^{9,d}$, S.~Ahmed$^{14}$, M.~Albrecht$^{4}$, A.~Amoroso$^{53A,53C}$, F.~F.~An$^{1}$, Q.~An$^{50,40}$, J.~Z.~Bai$^{1}$, Y.~Bai$^{39}$, O.~Bakina$^{24}$
% ...
   \\
   \vspace{0.2cm}
   (BESIII Collaboration)\\
   \vspace{0.2cm} {\it
$^{1}$ Institute of High Energy Physics, Beijing 100049, People's Republic of China\\
% ...
$^{4}$ Bochum Ruhr-University, D-44780 Bochum, Germany\\
% ...
$^{9}$ G.I. Budker Institute of Nuclear Physics SB RAS (BINP), Novosibirsk 630090, Russia\\
% ...
$^{14}$ Helmholtz Institute Mainz, Johann-Joachim-Becher-Weg 45, D-55099 Mainz, Germany\\
% ...
$^{24}$ Joint Institute for Nuclear Research, 141980 Dubna, Moscow region, Russia\\
% ...
$^{39}$ Southeast University, Nanjing 211100, People's Republic of China\\
$^{40}$ State Key Laboratory of Particle Detection and Electronics, Beijing 100049, Hefei 230026, People's Republic of China\\
% ...
$^{50}$ University of Science and Technology of China, Hefei 230026, People's Republic of China\\
% ...
$^{53}$ (A)University of Turin, I-10125, Turin, Italy; (B)University of Eastern Piedmont, I-15121, Alessandria, Italy; (C)INFN, I-10125, Turin, Italy\\
% ...
$^{d}$ Also at the Novosibirsk State University, Novosibirsk, 630090, Russia\\
% ...
    }
  \vspace{0.4cm}
}

\date{\today}
\begin{abstract}
  Using a data sample of $e^+e^-$ collision data corresponding to
  an integrated luminosity of 2.93~$\ifb$ collected with the BESIII
  detector at a center-of-mass energy of $\sqrt{s}= 3.773~\gev$,
  we search for the singly Cabibbo-suppressed decays $\dz\to\pipipi$,
  $\pipieta$, $\pietaeta$ and $\etaetaeta$ using the double tag method.
  The absolute branching fractions are measured to be $\BR(\dz\to\pipipi) = (2.0 \pm 0.4 \pm 0.3)\times 10^{-4}$, $\BR(\dz\to\pipieta) = (3.8 \pm 1.1 \pm 0.7)\times 10^{-4}$ and $\BR(\dz\to\pietaeta) = (7.3 \pm
  1.6 \pm 1.5)\times 10^{-4}$ with the statistical significances of $4.8\sigma$, $3.8\sigma$ and $5.5\sigma$, respectively, where the first uncertainties are statistical and the second ones systematic.
  No significant signal of $\dz\to\etaetaeta$ is found,
  and the upper limit on its decay branching fraction is set to be
  $\BR(\dz\to\etaetaeta) < 1.3 \times 10^{-4}$ at the 90\% confidence level.
\end{abstract}
\end{frontmatter}

\begin{multicols}{2}

\section{Introduction}
The study of charmed meson decays, which involve both strong and weak
interactions, is an interesting and challenging field in particle
physics.  Experimental measurements of charmed meson decays yield
essential information for understanding the intrinsic decay mechanism
and provide inputs to theoretical calculations and predictions.  For
example, Ref.~\cite{2008MGaspero} suggests that the measurement of the
branching fraction (BF) of the hadronic decay $\dz\to\pipipi$ may shed
light on the understanding of the role of isospin symmetry in $\dz$
decays to three-pion final states, and the isospin nature of the
non-resonant contribution.  Additionally, the study of the hadronic
decays of charmed mesons provides important inputs for the studies of
$B$ physics~\cite{2016LHCb}.

The singly Cabibbo-suppressed (SCS) decays of the $D^0$ meson to three
neutral pseudoscalar particles, $\dz\to\pipipi$, $\pipieta$,
$\pietaeta$ and $\etaetaeta$, proceed dominantly through internal
$W$-emission and $W$-exchange diagrams.
Experimental studies of these decays are challenging due to the
dominant presence of neutral particles (photons) in the final states,
low BFs and high backgrounds.  Until now, only a search for
$\dz\to\pipipi$ decay has been performed by the CLEO Collaboration
with a $\psi(3770)$ data sample of 281 $\ipb$ in
2006~\cite{2006Rubin}.  Using the ``single tag" (ST) method, in which
one $\dz$ or $\dbar$ meson is found in each event, they obtained a BF
upper limit of $3.5 \times 10^{-4}$ at the 90$\%$ confidence level
(C.L.).

In this Letter, we present measurements of the BFs of the SCS
decays $\dz\to\pipipi$, $\pipieta$, $\pietaeta$ and $\etaetaeta$ with
the ``double tag" (DT) technique and a data sample corresponding
to an integrated luminosity of 2.93~$\ifb$~\cite{2013Lumi}, collected
at a center-of-mass energy of $\sqrt{s} = 3.773$~GeV with the BESIII
detector at the BEPCII $\ee$ collider.  Throughout the Letter, charge
conjugate modes are always implied, unless explicitly mentioned.

\section{BESIII Detector and Monte Carlo Simulation}
BESIII~\cite{2009MAblikimDet} is a cylindrical spectrometer composed
of a helium-gas-based main drift chamber (MDC), a plastic scintillator
time-of-flight (TOF) system, a CsI(Tl) electromagnetic calorimeter
(EMC), a superconducting solenoid providing a 1.0~T magnetic field,
and a muon counter.  The charged particle momentum resolution in the
MDC is 0.5\% at a transverse momentum of 1 GeV/$c$ and the photon
energy resolution in the EMC at 1 GeV, is 2.5\% in the barrel region
and 5.0\% in the end-cap region.  Particle identification (PID)
combines the ionization energy loss ($dE/dx$) in the MDC with
information from the TOF
to identify particle types.  More details about the design and
performance of the detector are given in Ref.~\cite{2009MAblikimDet}.

GEANT4-based~\cite{GEANT} Monte Carlo (MC) simulation software is used
to understand the backgrounds and to determine the detection
efficiencies.  The generator
$\textsc{KKMC}$~\cite{2000SJadach,2001SJadach} is used to simulate the
$\ee$ collision incorporating the effects of beam-energy spread and
initial-state radiation (ISR).  An inclusive MC sample including
$\DDbar$, $\dpdm$ and non-$D\bar{D}$ events, ISR production of
$\psi(3686)$ and $J/\psi$, and continuum processes $\ee\to\qqbar$ ($q
= u, d, s$) is used to study the potential backgrounds.  The known
decay modes as specified in the Particle Data Group
(PDG)~\cite{2014PDG} are generated by
EVTGEN~\cite{2008RGPing,2001DJLange}, while the remaining unknown
decays of charmonium are modeled by LundCharm~\cite{2000JCChen}.

\section{Analysis Strategy}
At the $\psi(3770)$ resonance, $\dz\dbar$ pairs are produced in a
coherent $1^{--}$ state without additional particles.  A DT method,
which was first developed by the MARK-III
Collaboration~\cite{1986RMBaltrusaitis,1988JAdler}, is used to measure
the absolute BFs.  We first select ST events in which a $\dbar$ meson
is reconstructed in a specific hadronic decay mode.  Then we search
for $\dz$ decays in the remaining tracks, and DT events are those
where $\dz\dzbar$ pairs are fully reconstructed.  The absolute BFs for
$\dz$ decays are calculated by
 \begin{eqnarray}
 \BR^{\rm sig} &=& \frac{N^{\rm sig}_{\rm DT}}{\BR^{\rm int} \, \sum\limits_\alpha N^{\alpha}_{\rm ST} ~ {\rm \epsilon}^{\rm sig,\alpha}_{\rm DT}~/~{\rm \epsilon}^{\alpha}_{\rm ST}},
 \label{eq:absBr}
\end{eqnarray}
where the superscript `{\rm sig}' represents a specific $\dz$ signal
decay, $N^{\alpha}_{\rm ST}$, $\epsilon^{\alpha}_{\rm ST}$ and
$\epsilon^{\rm sig, \alpha}_{\rm DT}$ are the yield of ST events, the
ST detection efficiency and DT detection efficiency for a specific ST
mode $\alpha$, respectively, while $N^{\rm sig}_{\rm DT}$ is the total
yield for DT signal events, and $\BR^{\rm int}$ is the product of the
decay BFs for the intermediate states in the $\dz$ signal decay.

\section{Data Analysis}
Charged tracks are reconstructed from hits in the MDC and are required
to have a polar angle $\theta$ satisfying $|\cos\theta|<0.93$.  The
point of the closest approach of any charged track to the interaction
point (IP) is required to be within 1 cm in the plane perpendicular to
the beam and $\pm10$~cm along the beam.  Information from the TOF
system and the $dE/dx$ information in the MDC are combined to form PID
C.L.s for the $\pi$ and $K$ hypotheses.  Each track is assigned to the
particle type with the highest PID C.L.

Photon candidates are reconstructed using clusters of energy deposited in the
EMC crystals.  The energy is required to be larger than 25 MeV in the barrel
region ($|\cos\theta| < 0.8$) or 50 MeV in the end-cap region ($0.86 <
|\cos\theta| <0.92$).  The energy deposited in nearby TOF counters is
included to improve the reconstruction efficiency and energy
resolution.  The difference of the EMC time from the event start time
is required to be within $[0,700]$ ns to suppress electronic noise and
showers unrelated to the event.

The $\piz$ and $\eta$ candidates are reconstructed from photon pairs
by requiring the invariant masses $M_{\gamma\gamma}$ to satisfy $115 <
M_{\gamma\gamma} < 150~\mevcc$ or $515 < M_{\gamma\gamma} <
570~\mevcc$, respectively.  To improve the resolution, the photon
pairs are fitted kinematically constraining their masses to the
nominal $\piz$ or $\eta$ masses~\cite{2014PDG}, and the resulting
energies and momenta of the two photons are used for subsequent
analysis.

The ST candidates are selected by reconstructing $\dbar$ decays to
$K^{+}\pi^{-}, K^{+}\pi^{-}\pi^{0}$ and $K^{+}\pi^{-}\pi^{-}\pi^{+}$.
Two variables, the energy difference $\de \equiv E_D - E_{\rm beam}$
and the beam-energy-constrained mass $\mbc \equiv \sqrt{E_{\rm
    beam}^{2}/c^{4} - p^{2}_{D}/ c^{2}}$, are used to identify the
$\dbar$ candidates.  Here, $E_{\rm beam}$ is the beam energy, and
$E_{D}(p_{D})$ is the reconstructed energy (momentum) of the $\dbar$
candidate in the $e^+e^-$ center-of-mass system.  Those $\dbar$
candidates are accepted for further analysis that satisfy
$\mbc>1.83~\gevcc$ and mode-dependent $\de$
requirements, which are approximately three times the value of the
resolution around the $\dbar$ nominal mass~\cite{2014PDG}, as
summarized in Table~\ref{tab:stYield_Eff}.  For each ST
mode, if there is more than one candidate in the event, the one with
the minimum $|\de|$ is selected.

The $\mbc$ distributions of the accepted $\dbar$ candidates are
shown in Fig.~\ref{fig:stFit}, where $\dbar$ signals are observed
with relatively low backgrounds.  Binned
maximum likelihood fits to the $\mbc$ distributions are performed to
obtain the ST yields.  In the fits, the signal shape is modeled by the MC
simulated shape convolved with a Gaussian function representing the
difference between data and MC simulation coming from the beam-energy
spread, ISR, the $\psi(3770)$ line shape, and resolution.  The
combinatorial background is modeled by an ARGUS
function~\cite{1990argus}.  The ST yields are calculated by
subtracting the integrated ARGUS background yields from the total
events counted in the signal region $1.859 < \mbc < 1.871~\gevcc$.  The ST
efficiency is studied using the same procedure on the inclusive MC
sample.  The resulting ST yields and the corresponding ST efficiencies
are summarized in
Table~\ref{tab:stYield_Eff}.

\begin{table*}
  \begin{center}
  \footnotesize
  \caption{Requirements on  $\de$ (in $\gev$), ST yields in
    data ($N^{\alpha}_{\rm ST}$), ST ($\epsilon^{\alpha}_{\rm ST}$
    (in \%)) and DT ($\epsilon^{\pipipi, \alpha}_{\rm DT}$,
    $\epsilon^{\pipieta, \alpha}_{\rm DT}$, $\epsilon^{\pietaeta,
      \alpha}_{\rm DT}$ and $\epsilon^{\etaetaeta, \alpha}_{\rm DT}$
    (in \%)) efficiencies. The uncertainties are statistical only. BFs of
    $\piz$ and $\eta$ decays to two photons are not included in the
    efficiencies.}
  \begin{tabular}{c c c c}
    \hline \hline
    ST mode                                  & $K^{+}\pi^{-}$    & $K^{+}\pi^{-}\piz$ & $ K^{+}\pi^{-}\pi^{-}\pi^{+}$ \\ \hline
    $\de$                                    & $(-0.027,~0.025)$ & $(-0.071,~0.041)$  & $(-0.025,~0.022)$             \\
    $N^{\alpha}_{\rm ST}$                    & $530634 \pm  739$ & $1030144 \pm 1129$ & $707080 \pm 925$              \\
    $\epsilon^{\alpha}_{\rm ST}$             & $64.83 \pm 0.04$  & $33.75 \pm 0.02$   & $38.01 \pm 0.02$              \\
    $\epsilon^{\pipipi, \alpha}_{\rm DT}$    & $10.56\pm0.02$    & $4.46\pm 0.01$     & $4.78\pm 0.02$                \\
    $\epsilon^{\pipieta, \alpha}_{\rm DT}$   & $9.74\pm 0.02$    & $4.09\pm 0.01$     & $4.38\pm 0.01$                \\
    $\epsilon^{\pietaeta, \alpha}_{\rm DT}$  & $8.23\pm 0.02$    & $3.47\pm 0.01$     & $3.58\pm 0.01$                \\
    $\epsilon^{\etaetaeta, \alpha}_{\rm DT}$ & $10.02\pm 0.02$   & $4.14\pm 0.01$     & $4.57\pm 0.01$                \\
    \hline\hline
  \end{tabular}
  \label{tab:stYield_Eff}
  \end{center}
\end{table*}

\begin{figurehere}
  \centering
  \includegraphics[width=0.4\textwidth, height=0.6\textwidth]{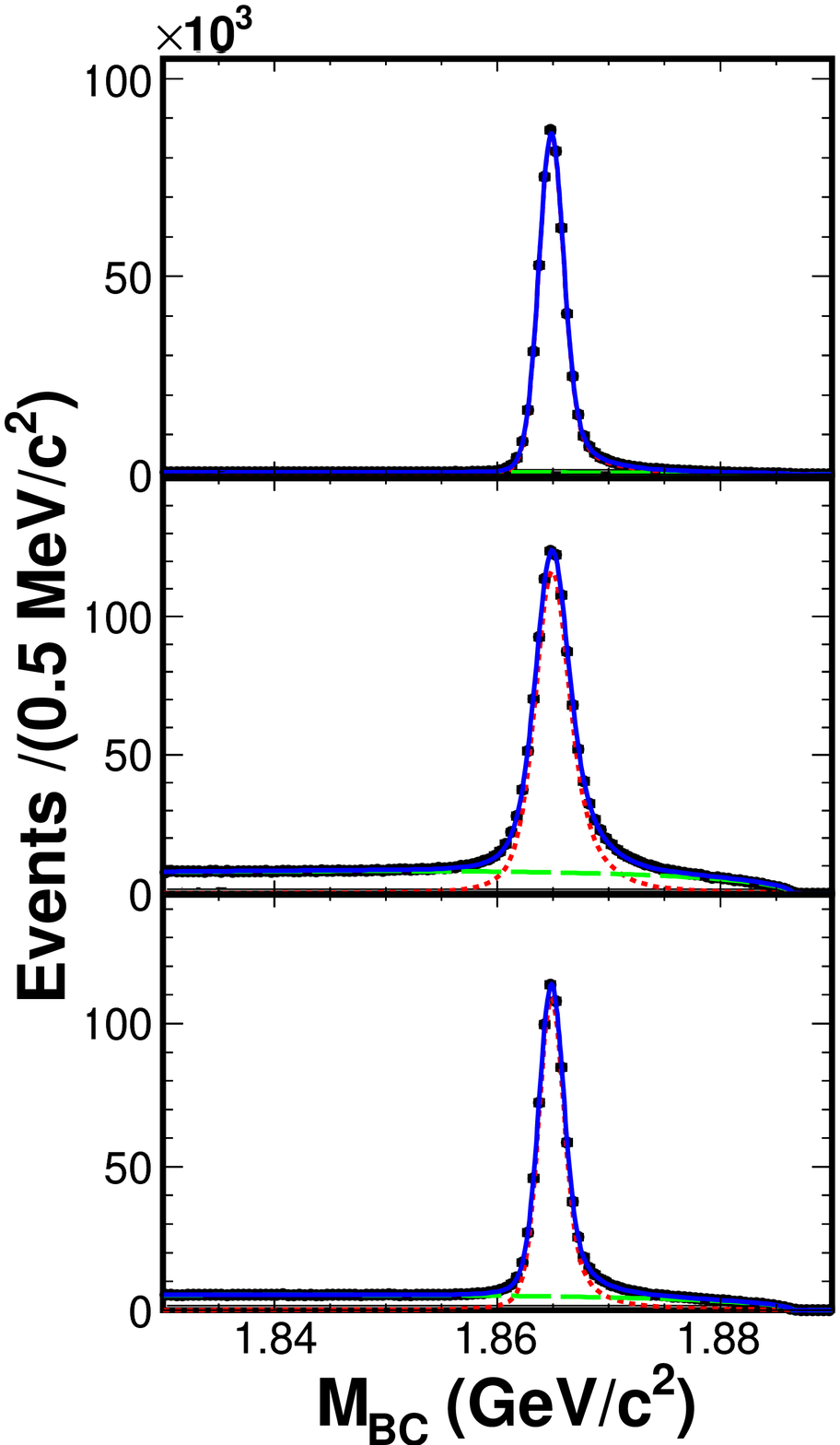}
  \put(-25,80){\bf (c)}
  \put(-25,165){\bf (b)}
  \put(-25,250){\bf (a)}
  \caption{(color online) Fits to the $\mbc$ distributions of the
    candidates for the ST modes: (a) $\dbar\to K^{+}\pi^{-}$, (b)
    $\dbar\to K^{+}\pi^{-}\piz$ and (c) $\dbar\to
    K^{+}\pi^{-}\pi^{-}\pi^{+}$. Points with an error bar are data,
    the blue solid lines are the total fit curves, the red dashed
    lines are the signal shapes, and the green long-dashed lines are
    the background shapes.  }
  \label{fig:stFit}
\end{figurehere}

Candidates for the SCS decays, $\dz\to\pipipi$, $\pipieta$,
$\pietaeta$ and $\etaetaeta$, are selected in the system recoiling
against the tagged $\dbar$.  Only events without any additional
charged track are chosen.  The $\dz$ signal decays are reconstructed
with any combination of the selected $\piz$ and $\eta$ candidates
that have not been used in the ST side and do not share the same
photon candidate.  To distinguish the signal decay from combinatorial
backgrounds, the energy difference $\Delta E$ and the beam-constrained
mass $M_{\rm BC}$ are also calculated for each accepted combination.
A $\dz$ candidate is accepted if it satisfies a mode-dependent $\de$
requirement, which corresponds to three times the value of the resolution
around the $\de$ peak based on MC simulation, as summarized in
Table~\ref{tab:summary}.  The shift and asymmetry of the $\de$
distributions are mainly due to the energy loss in the EMC for multi-photon final states.
If there are multiple combinations for a given signal decay in an
event, the one with the minimum $|\de|$ is selected.

Except for the decay $\dz\to\etaetaeta$, MC studies indicate that the
selected candidates have large backgrounds from
$\dz\to\piz\piz\piz\piz$ decay, which has a relatively large decay BF,
and contain some background events from cross feeds between signal channels.  Both
backgrounds peak around the nominal $\dz$ mass~\cite{2014PDG} in the
$\mbc$ distributions.  To reduce the background from
$\dz\to\piz\piz\piz\piz$ in $\dz\to\piz\piz\piz$ and $\piz\piz\eta$
decays, the joint chi-square
$\chi_{4\pi}^2=\sum_{i=1}^{4}\chi_{\pi_i}^2$ is required to be larger
than 20 if the candidate event has at least four independent $\piz$
candidates (not including $\piz$ candidates from the ST side).  Here,
$\chi_{\pi_i}=\frac{M_{\gamma\gamma}^i-m_{\piz}}{\sigma_{\gamma\gamma}^i}$
for the $i^{\rm th}$ $\piz$ candidate is calculated with the
$\gamma\gamma$ invariant mass $M_{\gamma\gamma}^i$ (before the kinematic
fit) and its resolution $\sigma_{\gamma\gamma}^i$, as well as the
$\piz$ nominal mass $m_{\piz}$~\cite{2014PDG}.  To reduce the cross
feed between the signal decays, we define the analogical joint
chi-square variables, $\chi^{2}_{ABC} =
(\frac{M_{\gamma\gamma}^1-m_{A}}{\sigma_{\gamma\gamma}^1})^{2} +
(\frac{M_{\gamma\gamma}^2-m_{B}}{\sigma_{\gamma\gamma}^2})^{2} +
(\frac{M_{\gamma\gamma}^3-m_{C}}{\sigma_{\gamma\gamma}^3})^{2}$, where
$m_{A(B,C)}$ is the nominal mass of $\piz$ or $\eta$~\cite{2014PDG},
and require $\chi^2_{\piz\piz\eta} >20$ for $\dz\to\pipipi$ decay,
$\chi^2_{\pipipi}>20$ for $\dz\to\pipieta$ decay as well as
$\chi^2_{\pipipi}>20$ and $\chi^2_{\piz\piz\eta}>20$ for
$\dz\to\pietaeta$ decay.

However, MC studies indicate that backgrounds remain from photon
mis-combinations in $\pi^0$ and $\eta$ candidates. These are
due to the matches of a good photon with noise in the EMC, which usually
corresponds to a fake low energy photon. Furthermore, the MC indicates
that this background can be reduced by requiring no other combination
with the same final state and with $\chi^2 < 20$.
For instance for $D^0 \to \pi^0 \pi^0 \pi^0$, this requirement loses
only 5\% of signal events while it rejects 30\% of mis-combination
background.

For $\dz\to\pipipi$ and $\pipieta$ decays, the events with any
$\piz\piz$ invariant mass satisfying $445 < M_{\piz\piz} < 535~\mevcc$
are vetoed to reject the backgrounds from the Cabibbo-favored (CF)
decays $\dz\to\ks\piz$ and $\ks\eta$ with $\ks\to\piz\piz$, which have
exactly the same final states as the signal channels.

With the above selection criteria, the $\mbc$ distributions of the
accepted $\dz$ candidate events in data are shown in
Fig.~\ref{fig:fitmbc}.  The $\dz\to\pipipi$, $\pipieta$ and
$\pietaeta$ signals are clear, but no obvious $\dz\to\etaetaeta$
signal is observed.  The peaking backgrounds are dominated by the
decay $\dz\to\piz\piz\piz\piz$, and the CF decays
$\dz\to\ks\piz/\eta$ for $\dz\to\pipipi/\eta$.  The contributions
from the cross feeds are small and will be considered in determining
the signal yields.  The mis-combination background is
negligible.

To determine the signal yields of the decays $\dz\to\pipipi$,
$\pipieta$, and $\pietaeta$, unbinned maximum likelihood fits are
performed to the $\mbc$ distributions.  The probability density
function (PDF) for signal is modeled with the MC simulated shape
convolved with a Gaussian function representing the resolution
difference and a potential mass shift between data and MC simulation.
The peaking backgrounds from the CF decay $\dz\to\ks\piz/\eta$ (BKG I)
and the decay $\dz\to\piz\piz\piz\piz$ (BKG II) as well as the cross feeds (BKG
III) are also included in the fit.  The combinatorial background (BKG
IV) is modeled by an ARGUS function~\cite{1990argus}.  The shapes of
the various peaking backgrounds are modeled with those of MC simulations,
and the corresponding magnitudes are fixed to the values estimated
with a data driven method.  We select a control sample of
$\dz\to\piz\piz\piz\piz$ from data with an approach similar to the
signal selection, and obtain the yield $N_{4\piz}$ from a fit to the
resulting $\mbc$ distribution.  A mixed MC sample, which includes the
possible resonant decays $\dz\to\bar{K}^{*}(892)^{0}\piz$, $\eta\piz$,
$\ks f'_{0}$, $f_{0}(980)\piz\piz$, $K^{0}_{L}\piz$, $\ks\ks$ and
$\eta'\piz$, is generated with known BFs~\cite{2014PDG} and is subject
to the selection criteria of $\dz\to\pipipi$ and
$\dz\to\piz\piz\piz\piz$ to evaluate the mis-identification rate
$\epsilon_{3\piz}$ and the detection efficiency $\epsilon_{4\piz}$,
respectively.  The magnitude of the background
$\dz\to\piz\piz\piz\piz$ in the selection of $\dz\to\pipipi$ is
given by $N_{4\piz}\cdot\epsilon_{3\piz}/\epsilon_{4\piz}$.
Similar data driven approaches are applied to determine the
magnitude of the peaking background $\dz\to\piz\piz\piz\piz$, the
cross feed and the number of CF decays $\dz\to\ks\piz/\eta$ in each signal
decay.  The resulting fits for $\dz\to\pipipi$, $\pipieta$ and
$\pietaeta$ are shown in Figs.~\ref{fig:fitmbc} (a), (b) and (c),
respectively.  The signal yields and statistical significances, which
are estimated from the likelihood difference between the fits with and
without the signal included after considering the change in the number
of degrees of freedom, are summarized in Table ~\ref{tab:summary}.

\begin{table*}
  \begin{center}
  \footnotesize
  \caption{Summary of $\de$ requirements, signal yields ($N_{\rm DT}^{\rm sig}$), statistical significances, BFs by this measurement and in the PDG~\cite{2014PDG}.
  The first and second uncertainties are statistical and systematic, respectively. The upper limit is set at the 90\% C.L..}
  \begin{tabular}{l c c c c c}
      \hline \hline
      Mode         & $\de~(\gev)$      & $N_{\rm DT}^{\rm sig}$     & Significance   & $\BR$~($\times 10^{-4}$)  & $\BR$$_{\rm PDG}$~($\times 10^{-4}$) \\ \hline
      $\pipipi$    & $(-0.115,~0.059)$ & $60\pm13$ & $4.8\sigma$    & $2.0 \pm 0.4 \pm 0.3$     & $<3.5$                               \\
      $\pipieta$   & $(-0.088,~0.053)$ & $42\pm12$ & $3.8\sigma$    & $3.8 \pm 1.1 \pm 0.7$     & $-$                                  \\
      $\pietaeta$  & $(-0.061,~0.045)$ & $27\pm6$  & $5.5\sigma$    & $7.3 \pm 1.6 \pm 1.5$     & $-$                                  \\
      $\etaetaeta$ & $(-0.030,~0.028)$ &  $-$      & $-$            & $< 1.3$                   & $-$                                  \\
      \hline\hline
  \end{tabular}
  \label{tab:summary}
  \end{center}
\end{table*}

\begin{figurehere}
  \centering
  \includegraphics[width=0.42\textwidth, height=0.80\textwidth]{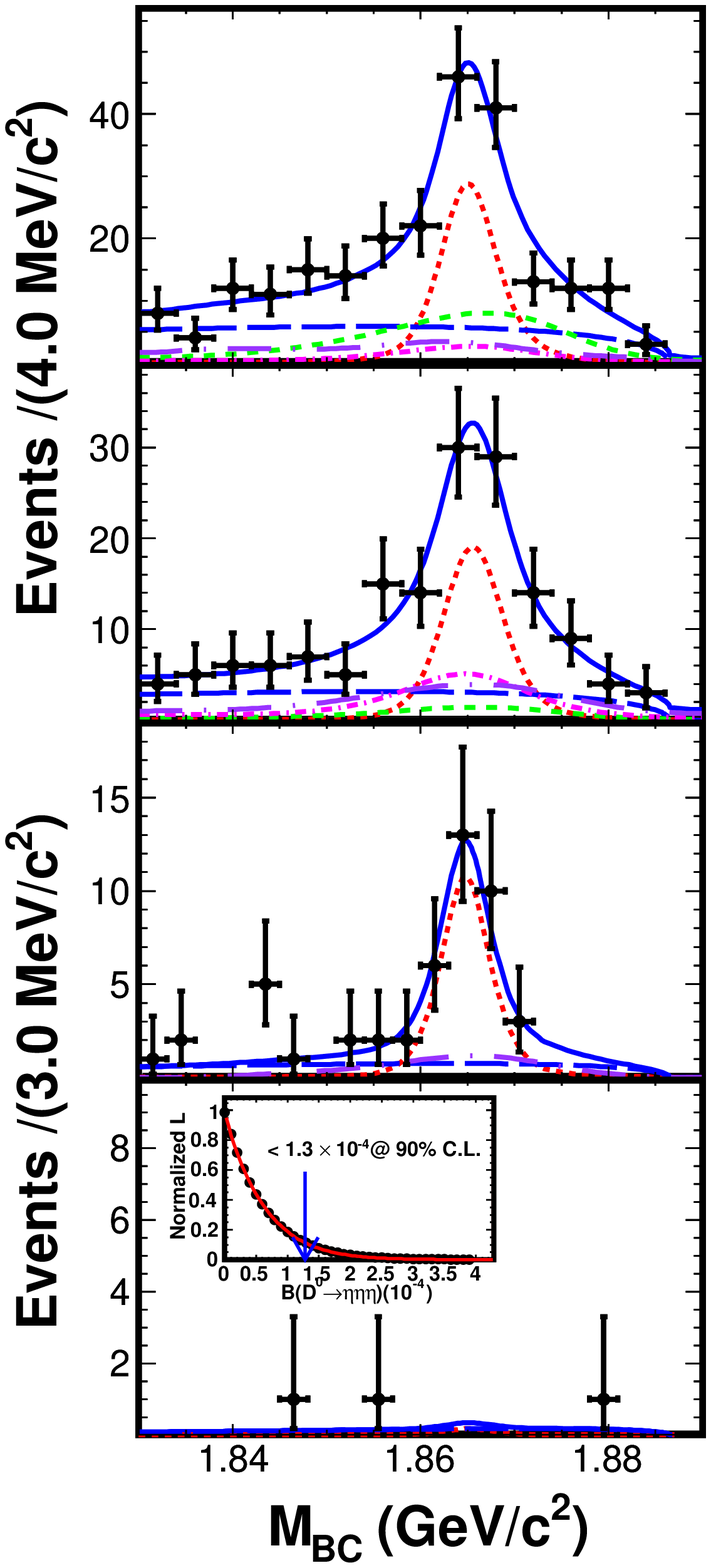}
  \put(-25,100){\bf (d)}
  \put(-25,185){\bf (c)}
  \put(-25,270){\bf (b)}
  \put(-25,355){\bf (a)}
  \caption{(color online) Fits to the $\mbc$ distributions of the
    accepted candidate events for (a) $\dz\to\pipipi$, (b)
    $\dz\to\pipieta$, (c) $\dz\to\pietaeta$ and (d)
    $\dz\to\etaetaeta$.  Dots with error bars are data, the blue solid
    lines are the total fit curves, and the red dotted lines are the
    signal shapes.  The green dashed, magenta dash-dotted, orange dash
    two-dotted and blue long-dashed lines denote BKG I, BKG II, BKG
    III and BKG IV (see text), respectively.  The violet long
    dash-dotted lines are the remaining $\dz\dbar$ background.  The
    inset in plot (d) shows the normalized likelihood distribution
    including the systematic uncertainty, as a function of the
    expected BF. The blue arrow indicates the upper limit on the BF at
    the 90\% C.L.  }
  \label{fig:fitmbc}
\end{figurehere}

Since no obvious $\dz\to\etaetaeta$ signal is observed, an upper limit
on its decay BF is determined.  We fit the $\mbc$ distribution of the
$\dz\to\etaetaeta$ candidate events, where the signal is described by
the MC simulated shape convoluted with a Gaussian function and the
background by an ARGUS function.  The parameters of the Gaussian
function are fixed to those obtained in the fit of $\dz\to\pietaeta$
decay.  The resultant best fit is shown in Fig.~\ref{fig:fitmbc} (d).
The PDF for the expected signal yield is taken to be the normalized
likelihood $\mathcal{L}$ versus the BF in the fit, incorporating the
systematic uncertainties as described below, and is shown as the inset
plot in Fig.~\ref{fig:fitmbc} (d).  The upper limit on the BF at the
90\% C.L., corresponding to $\int_{0}^{\rm up}
\mathcal{L}(x)dx/\int_{0}^{\infty} \mathcal{L}(x)dx=0.9$, is
calculated to be $<1.3\times10^{-4}$.

The detection efficiencies for various decays of interest must take
into account the effect of any intermediate states. The existence of
intermediate states in the $\dz$ three-body decays is investigated by
examining the corresponding Dalitz plots.  Except for the decay
$\dz\to\pietaeta$, no obvious intermediate states are observed.
Therefore, the detection efficiencies for the decays $\dz\to\pipipi$,
$\pipieta$ and $\etaetaeta$ are obtained with MC samples of three-body
phase space decay with uniform angular distributions.

For the decay $\dz\to\pietaeta$, the $a_{0}(980)^{0}$ is evident in
the $\piz\eta$ invariant mass $M_{\piz\eta}$ distribution.
Figure~\ref{fig:mpieta} shows the $M_{\piz\eta}$ spectrum of 23 events
with two entries per event from the data sample with additional
requirements $-0.023 < \de < 0.020~ \gev$ and $1.859 < \mbc < 1.871
~\gevcc$.  An unbinned maximum likelihood fit is performed on the
$M_{\piz\eta}$ distribution to determine the $a_{0}(980)^{0}$ signal
yield.

In the fit, the shape of the $\az$ is described with the shape from
the MC sample of $\dz\to a_{0}(980)^{0}\eta\to\pietaeta$, which has
two components: one with the $\piz$ combined with the correct $\eta$
coming from the $\az$ decay, and the other with the $\piz$ combined
with the wrong $\eta$ coming directly from the $\dz$ decay.  The first
peaks around the $\az$ mass, while the second contributes a broad
shape in the $M_{\piz\eta}$ distribution.  The MC shape is convolved
with a Gaussian function to account for the mass resolution difference
between data and MC simulation. In the MC simulation, the intermediate
$a_{0}(980)^{0}$ state is parameterized with the Flatt$\acute{\rm e}$
formula~\cite{1976Flatte} with the central mass and the $\az$ coupling
constants coming from the Crystal Barrel
experiment~\cite{2008ZouBS,1994ZouBS}.  The component from the direct
$\dz$ three-body decay is included in the fit, and its shape is the MC
simulated shape, which is similar to that of the wrong $\eta$
contribution in the $\az$ shape.  We also include the background in the
fit, where its shape is determined from the inclusive MC sample.  Both
magnitudes for the $\dz$ three-body decay component and background are
left free in the fit.  The fit curves are shown in
Fig.~\ref{fig:mpieta}.  The fit yields are $21\pm5$ events for the
$a_{0}(980)^{0}$ signal and $0\pm4$ events for the $\dz$ direct
three-body decay, which implies the predominant process in the
three-body decay of $\dz\to\pietaeta$ is $\dz\to a_{0}(980)^{0}\eta$.

We also perform a fit without the $a_{0}(980)^{0}$ signal included,
and the statistical significance of the $a_{0}(980)^{0}$ signal is
calculated with the change of likelihood value with respect to that of
the nominal fit taking into account the change of number of freedom in
the fit.  The significance for the $a_{0}(980)^{0}$ signal is only
2.6$\sigma$, although it is the predominant component in the
three-body decay.  Therefore, in the decay of $\dz\to\pietaeta$, the
detection efficiency is estimated with the MC sample of $\dz\to
a_{0}(980)^{0}\eta\to\pietaeta$ as described above.

The resultant DT efficiencies for various decays are listed in Table~\ref{tab:stYield_Eff}.
The BFs of these decays are calculated with Eq.~\ref{eq:absBr}, and summarized in Table~\ref{tab:summary}.

\begin{figurehere}
  \centering
  \includegraphics[width=0.40\textwidth, height=0.3\textwidth]{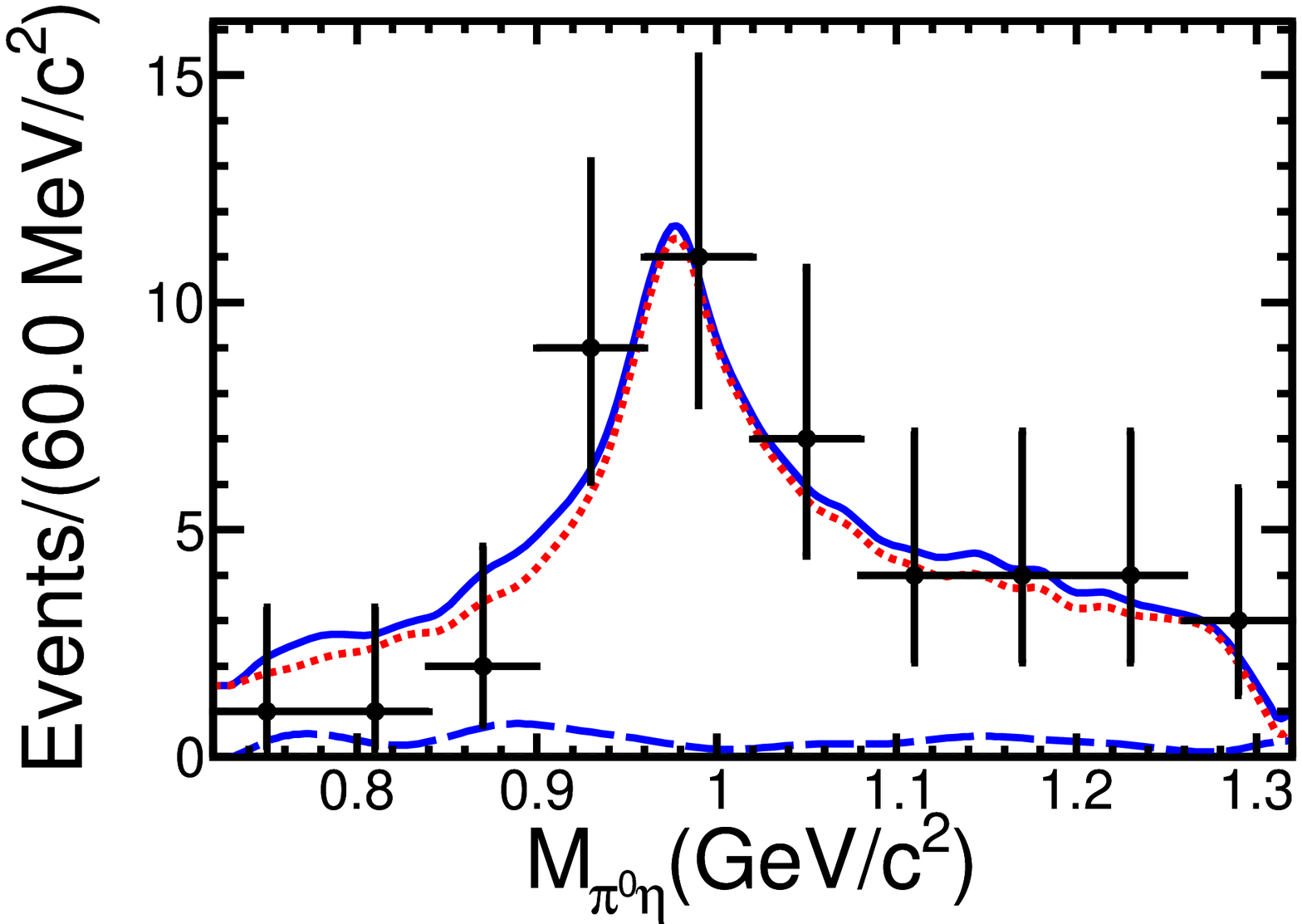}
  \caption{(color online) Fits to the $M_{\piz\eta}$ distribution.
    Dots with error bars are data, the blue solid line is the total
    fit curve, and the red dotted line is the signal shape.  The blue
    long-dashed line is the background estimated from the inclusive
    MC.  }
  \label{fig:mpieta}
\end{figurehere}

\section{Systematic Uncertainties}

With the DT technique, the BF measurements are insensitive to
systematics coming from the ST side since they mostly cancel.  For the
signal side, systematic uncertainties come mainly from the $\piz$ $(\eta)$
reconstruction efficiency, $\de$ resolution, CF background veto,
$\chi^{2}$ requirement, $\mbc$ fit, MC model, MC statistics, BFs of
$\piz$ and $\eta$ decays, and strong phase correction.

The $\piz$ reconstruction efficiency, including the photon detection
efficiency, is studied as a function of $\piz$ momentum using a
control sample of $\dz\to\kpipiz$ events.  The difference of the
$\piz$ reconstruction efficiencies between data and MC simulation is
regarded as the uncertainty related to $\piz$ reconstruction.  We
assume that the uncertainty due to reconstruction of the $\eta$ is the same as that for
the $\piz$.  The momentum weighted uncertainties of $\piz$ $(\eta)$
reconstruction efficiencies are taken as the associated systematic
uncertainties and are listed in Table~\ref{tab:uncertainties} for each
decay.

Uncertainty in the $\de$ resolution is studied by widening the $\de$
requirement from $3$ to $3.5$ times the resolution around the $\de$
peak.  For each decay, the resultant change of the BF is taken as the
systematic uncertainty.

To estimate the uncertainty due to the $\ks$ veto for $\dz\to\pipipi$
and $\pipieta$ decays, the measurement is repeated with an alternative
$\ks$ mass window rejection region of $450<M_{\piz\piz}<530~\mevcc$,
which is enlarged from $4.5\sigma$ to $5.0\sigma$ of the resolution.
The change of the BF for each decay is taken as the relevant
systematic uncertainty.

The uncertainty arising from the $\chi^{2}_{4\pi(ABC)}$ requirements
is investigated by repeating the measurement with an alternative
requirement $\chi^2_{4\pi(ABC)}<25$.  The resultant difference of the
BF is taken as the corresponding systematic uncertainty for each
decay.

Several aspects are considered to estimate the uncertainty related to
the $\mbc$ fit.  To examine the uncertainty in the fit range, a fit
with an alternative range of $(1.835,1.890)~\gevcc$ is performed.  The
uncertainty of the signal shape is examined with an alternative fit, in
which a Crystal Ball function is used to model the $\dz$ signal.  Due
to the long lifetime of $\ks$, the photons from $\piz$ (which are from
$\ks$ decay) decay do not originate from the IP.  To study the uncertainty due
to the imperfect simulation of the photon production vertex and its
abnormal incidence into the EMC, an alternative MC sample, in which
the $\ks$ lifetime is set to zero, is used to determine the
magnitude of BKG I.  The uncertainty in BKG II is investigated with an
alternative MC sample of $\dz\to\pipipi\piz$ generated as phase space
decay.  The uncertainty from BKG III is checked by varying its
magnitude by one standard deviation in the fit.  The uncertainty from
BKG IV is investigated by replacing the ARGUS function with the
inclusive MC simulated background shape.  For each of these sources,
the resultant difference of the signal yield is treated as the
corresponding systematic uncertainty for each decay.  The total
uncertainty associated with the $\mbc$ fit is the quadratic sum of the
above individual values.

The uncertainty in the MC model is examined by analyzing the
alternative MC events with and without involving the resonances
$f_{0}(980)$ and $a_{0}(980)^{0}$.  The maximum change in the
detection efficiency is taken as the systematic uncertainty.  For the
decay $\dz\to\pipipi$, the MC sample with $f_0(980)$ intermediate
state, $\dz\to f_0(980)\piz\to\pipipi$, is selected.  For the decay
$\dz\to\pipieta$, the MC samples with $f_0(980)$ or $a_0(980)^0$
intermediate states, $\dz\to
f_0(980)\eta\to\pipieta$ or $\dz\to a_0(980)^0\piz\to\pipieta$, are
chosen.  For the decay $\dz\to\pietaeta$, the MC sample of the direct
phase space decay is used.  As for the decay $\dz\to\etaetaeta$, no
uncertainty in the MC model is assigned due to the relatively small phase
space.

The uncertainty on the efficiency due to limited MC statistics is determined by
$\sqrt{\epsilon \, (1-\epsilon) / N}$. Here, $\epsilon$ is the detection
efficiency, and $N$ is the number of the generated MC events.  The
uncertainties of the BFs for $\piz$ and $\eta$ decays to two photons
are taken from the PDG~\cite{2014PDG}.

The uncertainty due to the quantum-correlation of the $\dz\dbar$ pair is
considered via the strong phase factor.  The absolute BF is calculated
by $\BR^{\rm sig}_{\rm CP\pm} = \frac{1}{1\mp C_{f}} \BR^{\rm sig}$,
where $\BR^{\rm sig}$ is calculated from Eq.~\ref{eq:absBr}, $C_{f}$
is the strong phase factor~\cite{2017DuanPF}, which is
$(-12.4\pm1.8)\%$, $(-8.7\pm1.6)\%$ and $(-7.0\pm1.3)\%$ for the ST
mode of $\dzbar\to K^{+}\pi^{-}$, $K^{+}\pi^{-}\piz$ and
$K^{+}\pi^{-}\pi^{-}\pi^{+}$, respectively.  The value of CP$+$ or CP$-$
that determine the largest difference in BFs is used to give the
systematic uncertainty.

Assuming all uncertainties, summarized in
Table~\ref{tab:uncertainties}, are independent, the total
uncertainties in the BF measurements are obtained by adding the
individual uncertainties in quadrature.

\section{Summary}
In summary, by analyzing an $\ee$ annihilation data sample of
$2.93~\ifb$ collected at $\sqrt{s} = 3.773~\gev$ with the BESIII
detector and using a DT method, we present the first observation of
the SCS decay $\dz\to\pietaeta$ with statistical significance of $5.5
\sigma$.  We find the first evidence for the SCS decays
$\dz\to\pipipi$ and $\pipieta$ with statistical significances of $4.8
\sigma$ and $3.8\sigma$, respectively.  The corresponding BFs are
measured to be $\BR(\dz\to\pipipi) = (2.0 \pm 0.4 \pm 0.3)\times
10^{-4}$, $\BR(\dz\to\pipieta) = (3.8 \pm 1.1 \pm 0.7)\times 10^{-4}$,
and $\BR(\dz\to\pietaeta) = (7.3 \pm 1.6 \pm 1.5)\times 10^{-4}$,
where the uncertainties are statistical and systematic, respectively.
We do not observe a $\dz\to\etaetaeta$ signal,
and the upper limit on its decay BF is
$\BR(\dz\to\etaetaeta)<1.3\times 10^{-4}$ at the 90\% C.L.  These
results are summarized in Table~\ref{tab:summary}, and the upper limit
in the PDG~\cite{2014PDG} is also listed.  The BF for $\dz\to\pipipi$
is consistent with the BF upper limit set by CLEO~\cite{2006Rubin} and is
approximately three times of its theoretical
prediction~\cite{2008MGasperoBABAR}, which indicates that the model
needs to be improved.

\begin{tablehere}
  \begin{center}
  \footnotesize
  \caption{Relative systematic uncertainties (in \%) in the BF measurements.}
  \begin{tabular}{l c c c c}
    \hline \hline
    Source                      & $\pipipi$ & $\pipieta$ & $\pietaeta$ & $\etaetaeta$       \\ \hline
    $\piz(\eta)$ reconstruction & 5.7       & 7.4        & 9.5         & 9.2                \\
    $\de$ requirement           & 0.9       & 1.9        & 1.5         & 1.5                \\
    CF background veto          & 0.6       & 0.8        & -           & -                  \\
    $\chi^{2}$ requirement      & 1.1       & 0.9        & 1.5         & -                  \\
    $\mbc$ fit                  & 5.0       & 7.6        & 5.2         & -                  \\
    MC model                    & 7.9       & 9.4        & 12.6        & -                  \\
    MC statistics               & 0.5       & 0.6        & 0.6         & 0.7                \\
    $\piz(\eta)$ BFs            & 0.1       & 0.5        & 1.0         & 1.5                \\
    Strong phase correction     & 10.5      & 9.4        & 10.7        & 10.7               \\ \hline
    Total                       & 15.3      & 17.2       & 20.0        & 14.3               \\
    \hline\hline
  \end{tabular}
  \label{tab:uncertainties}
  \end{center}
\end{tablehere}

\section{Acknowledgements}
The BESIII collaboration thanks the staff of BEPCII and the IHEP computing center for their strong support.
This work is supported in part by National Key Basic Research Program of China under Contract No. 2015CB856700; National Natural Science Foundation of China (NSFC) under Contracts Nos.
11335008, 11375170, 11425524, 11475164, 11475169, 11605196,11605198, 11625523, 11635010, 11705192, 11735014;
the Chinese Academy of Sciences (CAS) Large-Scale Scientific Facility Program;
the CAS Center for Excellence in Particle Physics (CCEPP);
Joint Large-Scale Scientific Facility Funds of the NSFC and CAS under Contracts Nos. U1532102, U1532257, U1532258, U1732263;
CAS Key Research Program of Frontier Sciences under Contracts Nos. QYZDJ-SSW-SLH003, QYZDJ-SSW-SLH040; 100 Talents Program of CAS; INPAC and Shanghai Key Laboratory for Particle Physics and Cosmology; German Research Foundation DFG under Contracts Nos. Collaborative Research Center CRC 1044, FOR 2359; Istituto Nazionale di Fisica Nucleare, Italy; Koninklijke Nederlandse Akademie van Wetenschappen (KNAW) under Contract No. 530-4CDP03; Ministry of Development of Turkey under Contract No. DPT2006K-120470; National Science and Technology fund; The Swedish Research Council; U. S. Department of Energy under Contracts Nos. DE-FG02-05ER41374, DE-SC-0010118, DE-SC-0010504, DE-SC-0012069; University of Groningen (RuG) and the Helmholtzzentrum fuer Schwerionenforschung GmbH (GSI), Darmstadt.

\end{multicols}
\end{document}